\documentclass[12pt]{article}
\usepackage{amssymb}
\def\be{\begin{equation}}
\def\ee{\end{equation}}
\def\bea{\begin{eqnarray}}
\def\eea{\end{eqnarray}}
\def\p{\partial}

\def\half{{\textstyle{1\over2}}}

\def\pl#1{{\sl Phys.~Lett.~\bf B#1}}
\def\pr#1{{\sl Phys.~Rev.~\bf D#1}}
\def\prl#1{{\sl Phys.~Rev. Lett.~\bf #1}}

\def\cqg#1{{\sl Class.~Quant.~Grav.~\bf #1}}

\catcode`\@=11
\def\@citex[#1]#2{%
\if@filesw \immediate \write \@auxout {\string \citation {#2}}\fi
\@tempcntb\m@ne \let\@h@ld\relax \def\@citea{}%
\@cite{%
  \@for \@citeb:=#2\do {%
    \@ifundefined {b@\@citeb}%
      {\@h@ld\@citea\@tempcntb\m@ne{\bf ?}%
      \@warning {Citation `\@citeb ' on page \thepage \space undefined}}%
      {\@tempcnta\@tempcntb \advance\@tempcnta\@ne%
      \@tempcntb\number\csname b@\@citeb \endcsname \relax%
      \ifnum\@tempcnta=\@tempcntb 
        \ifx\@h@ld\relax%
          \edef \@h@ld{\@citea\csname b@\@citeb\endcsname}%
        \else%
          \edef\@h@ld{\ifmmode{-}\else--\fi\csname b@\@citeb\endcsname}%
        \fi%
      \else
        \@h@ld\@citea\csname b@\@citeb \endcsname%
        \let\@h@ld\relax%
      \fi}%
    \def\@citea{,\penalty\@highpenalty\,}%
  }\@h@ld
}{#1}}

\def\@citeb#1#2{{[#1]\if@tempswa , #2\fi}}
\def\@citeu#1#2{{$^{#1}$\if@tempswa , #2\fi }}
\def\@citep#1#2{{#1\if@tempswa , #2\fi}}

\def\bcites{         
        \catcode`\@=11
        \let\@cite=\@citeb
        \catcode`\@=12
}

\def\upcites{         
        \catcode`\@=11
        \let\@cite=\@citeu
        \catcode`\@=12
}

\def\plaincites{      
        \catcode`\@=11
        \let\@cite=\@citep
        \catcode`\@=12
}

\newcount\hour
\newcount\minute
\newtoks\amorpm
\hour=\time\divide\hour by 60
\minute=\time{\multiply\hour by 60 \global\advance\minute by-\hour}
\edef\standardtime{{\ifnum\hour<12 \global\amorpm={am}%
        \else\global\amorpm={pm}\advance\hour by-12 \fi
        \ifnum\hour=0 \hour=12 \fi
        \number\hour:\ifnum\minute<10 0\fi\number\minute\the\amorpm}}
\edef\militarytime{\number\hour:\ifnum\minute<10 0\fi\number\minute}

\def\draftlabel#1{{\@bsphack\if@filesw {\let\thepage\relax
   \xdef\@gtempa{\write\@auxout{\string
      \newlabel{#1}{{\@currentlabel}{\thepage}}}}}\@gtempa
   \if@nobreak \ifvmode\nobreak\fi\fi\fi\@esphack}
        \gdef\@eqnlabel{#1}}
\def\@eqnlabel{}
\def\@vacuum{}
\def\marginnote#1{}
\def\draftmarginnote#1{\marginpar{\raggedright\scriptsize\tt#1}}
\def\draft{
        \pagestyle{plain}
        \overfullrule=2pt
        \oddsidemargin -.5truein
        \def\@oddhead{\sl \phantom{\today\quad\militarytime} \hfil
        \smash{\Large\sl DRAFT} \hfil \today\quad\militarytime}
        \let\@evenhead\@oddhead
        \let\label=\draftlabel
        \let\marginnote=\draftmarginnote
        \def\ps@empty{\let\@mkboth\@gobbletwo
        \def\@oddfoot{\hfil \smash{\Large\sl DRAFT} \hfil}
        \let\@evenfoot\@oddhead}
        \def\@eqnnum{(\theequation)\rlap{\kern\marginparsep\tt\@eqnlabel}%
        \global\let\@eqnlabel\@vacuum}  }

\begin{document}

\hfill UTHET-04-0101

\vspace{-0.2cm}

\begin{center}
\Large
{\bf Large mass expansion of quasi-normal modes in AdS$_5$}
\normalsize

\vspace{0.8cm}
{\bf George Siopsis}
\footnote{
Research supported by the US Department of Energy under grant
DE-FG05-91ER40627.}
\\ Department of Physics
and Astronomy, \\
The University of Tennessee, Knoxville, \\
TN 37996 - 1200, USA. \\
{\tt email: gsiopsis@utk.edu}

\end{center}

\vspace{0.8cm}
\large
\centerline{\bf Abstract}
\normalsize
\vspace{.5cm}

We calculate analytically
the asymptotic form of quasi-normal frequencies for massive scalar perturbations of large black holes in
AdS$_5$.
We solve the wave equation, which reduces to a Heun equation, perturbatively
and calculate the wave function as an expansion in $1/m$, where $m$ is the mass of
the mode.
The zeroth-order results agree with expressions derived by numerical analysis.
We also calculate the first-order corrections to frequencies explicitly
and show that they are of order
$1/m$.

\newpage

Quasi-normal modes of black holes in asymptotically AdS space-times have received
a lot of attention recently~\cite{bibq1,bibq2,bibq3,bibq4,bibq5,bibq6,bibq7,bibq8,bibq9,bibq10,bibq11,bibq12,bibq13,bibw1,bibw2,bibw3,bibw4}
and should shed some light on the AdS/CFT correspondence.
They are derived as complex eigenvalues corresponding to a wave equation which
is solved subject to the
conditions that the flux be ingoing at the horizon and the wave-function vanish
at the boundary of AdS space.
The wave equation reduces to a Hypergeometric equation in three dimensions (AdS$_3$)
and can therefore be solved exactly~\cite{bibq7,bibq13}.
In higher dimensions, an analytic solution is not readily available, as the
wave equation develops unphysical singularities.
It has been possible to obtain numerical values for quasi-normal frequencies~\cite{bibq2,bibq14,bibr1,bibr2}.

More recently, we proposed an analytic method of calculating
massless quasi-normal modes of
large AdS black holes. We applied the method to the calculation of low-lying
frequencies~\cite{bibus} as well as in the asymptotic regime~\cite{bibus2} in AdS$_5$. The method was based on a
perturbative expansion
of the wave equation, which reduced to the Heun equation~\cite{bibq15}, in the dimensionless parameter $\omega / T_H$, where $\omega$
is the frequency of the mode and $T_H$ is the (high) Hawking temperature of the black hole.
This is a non-trivial expansion, for the dependence of the wave-function on $\omega / T_H$ changes as one moves from the asymptotic boundary of AdS space to the horizon
of the black hole.
We showed that our results were in agreement with numerical results~\cite{bibq2,bibq14}.

Here we extend the discussion of~\cite{bibus,bibus2} to the case of massive scalar modes.
The quasi-normal frequencies determine the poles of the retarded correlation
functions of dual operators in finite-temperature $\mathcal{N} = 4~SU(N)$ SYM theory in the large-$N$, large 't Hooft coupling limit~\cite{bibns}.
They have also been recently studied in~\cite{bibother} where they were shown to arise in complexified geodesics.
We shall derive a systematic analytic expansion of the frequencies in powers of $1/m$,
where $m$ is the mass of the mode.
Numerical results indicate~\cite{bibns} that in the limit of large mass,
\be\label{eqA} \frac{\omega_n}{T_H} \sim 2\pi (\pm 1 -i) (n+h_+ - {\textstyle{\frac{3}{2}}})
\ \ , \ \ n = 1,2,\dots \ee
where $h_\pm = 1\pm \sqrt{1+ m^2R^2}$.
Our results confirm analytically the above asymptotic result of numerical analysis.
Our method is an extension of the discussion in~\cite{bibus2} where the numerical result~(\ref{eqA}) was confirmed analytically in the massless case ($m=0$).
We also calculate the first-order correction and show that it is of order $1/h_+$.
A similar procedure was followed in~\cite{bibus}.


The metric of a five-dimensional AdS black hole may be written as
\be
ds^2 = -f(r)\, dt^2 + \frac{dr^2}{ f(r)} +r^2 d \Omega_3^2
\ \ , \ \
f(r) = \frac{r^2}{R^2} +1 - \frac{\omega_4 M}{r^2}\ee
where $R$ is the AdS radius and $M$ is the mass of the black hole.
For a large black hole, the metric simplifies to
\be
ds^2 = -\hat f(r)\,  dt^2 + \frac{dr^2}{\hat f(r)} +r^2 ds^2 (\mathbb{E}^3)
\ \ , \ \
\hat f(r) = \frac{r^2}{R^2} - \frac{\omega_4 M}{r^2}\ee
The Hawking temperature is
\be
T_H = \frac{r_h}{\pi R^2}
\ee
where $r_h$ is the radius of the horizon,
\be
r_h = R\; \left[ \frac{\omega_4 M}{R^2} \right] ^{1/4}
\ee
The wave equation for a massive scalar of mass $m$ is
\be\label{eq5}
\frac{1}{r^3}\p_r (r^5\, h(r)\, \p_r \Phi) -\frac{R^4}{ r^2\, h(r) }\p_{t}^2\Phi - \frac{R^2}{r^2}\; \vec\nabla^2\Phi = m^2 \Phi
\ \ , \ \
h(r) = 1- \frac{r_h^4}{r^4}
\ee
We are interested in solving this equation for
a wave which is ingoing at the horizon and vanishes at infinity. These boundary
conditions yield a discrete set of complex frequencies (quasi-normal modes).
The solution may be written as
\be 
\Phi = e^{i(\omega t - \vec p\cdot \vec x)} \Psi (r)
\ee
Upon changing the coordinate $r$ to $y$,
\be
y = \frac{r^2}{r_h^2} 
\ee
the wave equation becomes
\be\label{eq24}
\left( y(y^2-1) \Psi' \right)' + \left(\frac{\hat\omega^2}{4}\, \frac{y^2}{y^2-1} - \frac{\hat p^2}{4} - \hat m^2 y\right)\Psi = 0
\ee
where we have introduced the dimensionless variables
\be
\hat\omega = \frac{\omega R^2}{r_h} = \frac{\omega}{\pi T_H}, \ \ \ \ \
\hat p = \frac{|\vec p|R}{r_h} = \frac{|\vec p|}{\pi R T_H}
, \ \ \ \ \ \hat m = mR
\ee
Two independent solutions are obtained by examining the behavior near the
horizon ($y\to 1$),
\be\label{eqn1} \Psi_\pm \sim (y-1)^{\pm i\hat\omega/4}\ee
where $\Psi_+$ is outgoing and $\Psi_-$ is ingoing.
We ought to choose $\Psi_-$ for quasi-normal modes.

A different set of linearly independent solutions is obtained by studying the
behavior at large $r$
($y\to \infty$). We obtain
\be\label{eq27} \Psi\sim y^{-h_\pm} \ \ , \ \ h_\pm = 1\pm \sqrt{1+\hat m^2}\ee
For quasi-normal modes, we are interested
in the solution which vanishes at the boundary,
hence it behaves as $\Psi\sim y^{-h_+}$ as $y\to\infty$.
Combining this with the requirement that $\Psi = \Psi_-$ (eq.~(\ref{eqn1}))
leads to a discrete spectrum of quasi-normal frequencies.

By considering the other (unphysical) singularity at $y=-1$, we obtain
yet another set of linearly independent wave-functions behaving as
\be \Psi \sim (y+1)^{\pm \hat\omega /4} \ee
near $y=-1$. There is no restriction on the behavior of the wave-function at this singularity.
Nevertheless, as we shall see, it will be advantageous to isolate the behavior
near $y=-1$.

Isolating the behavior at the two singularities $y=\pm 1$, we shall
write the wave-function as
\be\label{eq25}
\Psi (y) = (y-1)^{-i\hat\omega/4} (y+1)^{-\hat\omega/4} F(y)
\ee
The selection of the exponent at the $y=-1$ singularity is arbitrary and not
guided by a physical principle. We shall show that it leads to a convenient
perturbative calculation of {\em half} of the modes.
Selecting the other exponent, $+\hat\omega/4$, similarly yields the other set
of modes without additional computational difficulties. The latter have the
same imaginary part as the former, but opposite real parts, as will become evident.

It is easily deduced from eqs.~(\ref{eq24}) and (\ref{eq25}) that
the function $F(y)$ satisfies the Heun equation
\bea
y(y^2-1) F'' + \left\{ \left( 3- \frac{1+i}{2}\, \hat\omega \right) y^2 + \frac{1-i}{2}\, \hat\omega y -1 \right\} F' & & \nonumber \\
+ \left\{ \frac{\hat\omega}{2}\left( \frac{i\hat\omega}{4} - 1-i\right) y -\hat m^2 y + (1-i)\frac{\hat\omega}{4} - \frac{\hat p^2}{4} \right\}\; F &=& 0 \label{w1}
\eea
We wish to solve this equation in a region in the complex $y$-plane
containing $|y|\ge 1$, which includes the
physical regime $r> r_h$.
For large $\hat\omega$, the constant terms in the respective Polynomial coefficients of $F'$ and $F$ in~(\ref{w1}) are small compared with the other terms, so they may be dropped.
Eq.~(\ref{w1}) may then be approximated by the Hypergeometric equation
\be\label{w2x}
(y^2-1) F'' + \left\{ \left( 3- \frac{1+i}{2}\, \hat\omega \right) y + \frac{1-i}{2}\, \hat\omega \right\} F'
+ \left\{ \frac{\hat\omega}{2}\left( \frac{i\hat\omega}{4} - 1-i\right)
- \hat m^2 \right\} \; F =0 \label{w2}
\ee
in the asymptotic limit of large frequencies $\hat\omega$.
Two linearly independent solutions of (\ref{w2x}) are
\be\label{eqhsol} \mathcal{F}_1 = F(a_+,a_-;c;-x) \ \ , \ \ \mathcal{F}_2 =
x^{1-c} F(1+a_+-c, 1+a_--c;2-c;-x)\ee
where
\be\label{eqxy} a_\pm = h_\pm -{\textstyle{\frac{1+i}{4}}}\,\hat\omega
\quad,\quad c = {\textstyle{\frac{3}{2}}} - {\textstyle{\frac{i}{2}}}\,\hat\omega \ \ , \ \ x = \frac{y-1}{2}\ee
The solution of~(\ref{w2x}) which is regular at the horizon ($x=0$) is the Hypergeometric function $\mathcal{F}_1$ (eq.~(\ref{eqhsol})).
Matching the scaling behavior~(\ref{eq27}) at the boundary of AdS$_5$, we obtain
a different set of linearly independent solutions of~(\ref{w2}),
\be\label{eqsolinf} \mathcal{K}_\pm = (x+1)^{-a_\pm} F(a_\pm, c-a_\mp ; a_\pm -a_\mp +1; 1/(x+1))\ee
We have $\mathcal{K}_\pm\sim x^{-a_\pm}$ as $x\to\infty$, and correspondingly,
$\Psi\sim x^{-h_\pm}$ (from eq.~(\ref{eq25}) using~(\ref{eqxy})).
Therefore,
our choice is $\mathcal{K}_+$, since it leads to $\Psi\to 0$ as $x\to\infty$.
It is a linear combination of $\mathcal{F}_1$ and $\mathcal{F}_2$ (eq.~(\ref{eqhsol})). We easily obtain from standard Hypergeometric identities,
\be \mathcal{K}_+ = \mathcal{A}_0 \mathcal{F}_1 + \mathcal{B}_0
\mathcal{F}_2\ee
where
\be\label{eq21} \mathcal{A}_0 = \frac{\Gamma(1-c)\Gamma(1-a_-+a_+)}{\Gamma(1-a_-)\Gamma(1-c+a_+)}
\ \ , \ \ \mathcal{B}_0 = \frac{\Gamma(c-1)\Gamma(1+a_+-a_-)}{\Gamma(a_+)\Gamma(c-a_-)}\ee
For the correct behavior at the horizon, we demand
\be \mathcal{B}_0 = 0\ee
which leads to two different conditions,
\be\label{eqqnf} c-a_- = -n+1 \ \ , \ \ n = 1,2,\dots\ee
or
\be\label{eqqnf2} a_+ = -n+1 \ \ , \ \ n = 1,2,\dots\ee
We obtain a set of quasi-normal frequencies by solving~(\ref{eqqnf}),
\be\label{eqo} \hat\omega_n = -2(1+ i)(n+h_+-{\textstyle{\frac{3}{2}}}) \ \ , \ \ n=1,2,\dots\ee
Another set is obtained from~(\ref{eqqnf2}),
\be \hat\omega_n = 2(1- i)(n+h_+-1) \ \ , \ \ n=1,2,\dots\ee
To leading order, the two sets of modes have the same imaginary parts but opposite real parts ($\hat\omega_n \sim 2(\pm 1 - i) (n+h_+)$). This is
in agreement with numerical results~\cite{bibq14} (see eq.~(\ref{eqA})).
To go beyond the zeroth-order approximation, it is computationally advantageous
to work with the first set of modes~(\ref{eqo}), which is a consequence of~(\ref{eqqnf}).
The other set of modes, with positive real part, can be obtained without additional effort by choosing the exponent $+\hat\omega/4$ at the unphysical singularity ($y=-1$), instead (eq.~(\ref{eq25})).

The above approximation $\mathcal{K}_+$~(eq.~(\ref{eqsolinf})) to the exact wave-function satisfying the
Heun equation~(\ref{w2}) may be used as the basis for a systematic calculation
of corrections to the asymptotic form~(\ref{eqo}) of quasi-normal frequencies.
To this end,
let us write the Heun equation~(\ref{w2}) as
\be\label{w3} (\mathcal{H}_0 +\mathcal{H}_1) F = 0\ee
where
\be \mathcal{H}_0  = (1-y^2)\frac{d^2}{dy^2} +
\left\{ \frac{i-1}{2}\, \hat\omega
- \left( 3- \frac{i+ 1}{2}\, \hat\omega \right) y \right\}
\frac{d}{dy}
- \frac{\hat\omega}{2}\left( \frac{i\hat\omega}{4} - 1-i\right)
-\hat m^2
\ee
and
\be\mathcal{H}_1 = \frac{1}{y} \left(\frac{d}{dy} + (i- 1)\frac{\hat\omega}{4} + \frac{\hat p^2}{4}
\right)\ee
The zeroth-order equation,
\be\label{eqzero} \mathcal{H}_0 F_0 = 0\ee
is the approximation we discussed above: the Hypergeometric eq.~(\ref{w2})
whose solution is
\be F_0 = \mathcal{K}_+\ee
(eq.~(\ref{eqsolinf})).
By treating $\mathcal{H}_1$ as a perturbation, we may
expand the wave-function,
\be F = F_0 + F_1 + \dots\ee
and solve eq.~(\ref{w3}) perturbatively. Corrections to the quasi-normal
frequencies~(\ref{eqo}) may then be obtained once an explicit expression
for the correction to $F_0$ has been calculated.

We shall restrict attention to the vanishing momentum case, $\hat p = 0$.
Extension to the more general case is straightforward.
We have
\be\label{eq33} F_1(x) = \mathcal{K}_-(x) \int_x^\infty \frac{\mathcal{K}_+ \mathcal{H}_1 F_0}{\mathcal{W}}
- \mathcal{K}_+(x) \int_x^\infty \frac{\mathcal{K}_- \mathcal{H}_1 F_0 }{\mathcal{W}}
\ee
where $\mathcal{K}_\pm$ are the two linearly independent solutions~(\ref{eqsolinf})
of eq.~(\ref{eqzero}) and $\mathcal{W}$ is their Wronskian,
\be \mathcal{W} = (a_+-a_-) x^{-c} (x+1)^{c-a_+-a_--1}
\ee
To study the behavior near the horizon ($x\to 0$), we may first analytically
continue the parameters so that the integrals in~(\ref{eq33}) remain finite
as $x\to 0$. Alternatively, we may introduce a cutoff $\xi$ and notice
that the part which is cut,
\be \delta F_1(x) = \mathcal{K}_-(x) \int_x^\xi \frac{\mathcal{K}_+ \mathcal{H}_1 F_0}{\mathcal{W}}
- \mathcal{K}_+(x) \int_x^\xi \frac{\mathcal{K}_- \mathcal{H}_1 F_0 }{\mathcal{W}}
\ee
remains finite as $x\to 0$. This is most easily seen by replacing the set of
solutions $\mathcal{K}_\pm$ with $\mathcal{F}_1$ and $\mathcal{F}_2$ (eq.~(\ref{eqhsol})). Therefore, the analytic continuation of the parameters does not
affect the singularity at the horizon. Writing
\be F_1(x) \sim \mathcal{A}_1 + \mathcal{B}_1 x^{1-c}\ \  \ \ (x\to 0)\ee
we therefore deduce
\be
\mathcal{B}_1 = \beta_- \int_0^\infty \frac{\mathcal{K}_+\mathcal{H}_1 F_0}{\mathcal{W}}
- \beta_+ \int_0^\infty \frac{\mathcal{K}_-\mathcal{H}_1 F_0}{\mathcal{W}}
\ee
where
\be\beta_\pm = \frac{\Gamma(c-1)\Gamma(1+a_\pm -a_\mp)}{\Gamma(a_\pm)\Gamma(c-a_\mp)}\ee
On account of~(\ref{eqqnf}), we have $\beta_+ = 0$, therefore,
\be\label{eq39}
\mathcal{B}_1 = \beta_- \int_0^\infty \frac{\mathcal{K}_+\mathcal{H}_1 F_0}{\mathcal{W}}
\ee
Hence to first order, the quasi-normal frequencies are obtained as solutions of
\be\label{eq40} \mathcal{B}_0 + \mathcal{B}_1 = 0\ee
where $\mathcal{B}_0$ is given by~(\ref{eq21}).

Let us first solve the $n=1$ case in order to calculate the first-order correction to $\hat\omega_1$.
In this case $a_-=c$ (eq.~(\ref{eqqnf})), and so
\be \mathcal{K}_+ = F_0 = (x+1)^{-a_+}
\ee
Also, $\beta_- = \frac{a_--a_+}{a_--1}$, and
\be \mathcal{H}_1 F_0 = \frac{- a_+ (x+1)^{-1} +a_+-a_- - 1}{2(2x+1)}
\ F_0\ee
It follows that
\be
\mathcal{B}_1 = \frac{1}{a_--1} \int_0^\infty dx x^{a_-} (x+1)^{1-a_+}\frac{- a_+ (x+1)^{-1} +a_+-a_--1 }{2(2x+1)}
\ee
Using
\be \int_0^\infty dx \frac{x^\lambda (1+x)^{-\nu}}{ 2x+1} = B(1+\lambda, \nu-\lambda) F(1, 1+\lambda; 1+\nu;-1)\ee
we deduce
\bea
\mathcal{B}_1 &=& \frac{\Gamma(a_--1)\Gamma(1+a_+-a_-)a_-}{2\Gamma(a_+)(a_+-a_-)}
\ \left\{ F(1,1+a_-;1+a_+;-1) - F(1,1+a_-;a_+;-1)\right\}\nonumber\\
&=& B(a_--1,1+a_+-a_-)\ \frac{a_-^2}{2(a_+-a_-)a_+(1+a_+)} \
F(2,a_-+1;a_++2;-1)\nonumber\\
&=& B(a_--1,1+a_+-a_-)\ \left\{ \frac{i}{4h_+} + \dots\right\}
\eea
where we used $a_\pm = (i\pm 1) h_+ + \dots$ and expanded
\be\label{eqhyp} F(1,\alpha;\beta; z)
= \left(1-\frac{\alpha}{\beta}\ z\right)^{-1}+ \left(\frac{1}{\alpha} - \frac{1}{\beta} \right)
\frac{\alpha^2 z^2}{\beta^2} \left(1-\frac{\alpha}{\beta}\ z\right)^{-3} + \dots\ee
for large $\alpha$ and $\beta$.

Setting
\be c-a_- = \epsilon\ee
we obtain from eq.~(\ref{eq21})
\be \mathcal{B}_0 = \epsilon B(a_--1,1+a_+-a_-) + \dots\ee
Therefore, demanding $\mathcal{B}_0 + \mathcal{B}_1 = 0$ (eq.~(\ref{eq40})), we obtain the first-order correction
\be \epsilon = -\frac{i}{4h_+}\ee
and the corresponding frequency
\be\label{eqoe1} \hat\omega_1 = -2(1+i)\left( h_+-\half + \frac{i}{4h_+}\right)\ee
Next, we discuss the case $n=2$. We have
\be\mathcal{K}_+ = F_0(x) = (x+1)^{-a_+} \left\{ 1 - \frac{a_+}{a_+-a_-+1} (x+1)^{-1} \right\}\ee
Working as before, we deduce
\be \mathcal{B}_1 = \frac{1}{2(a_--1)(a_--2)}\int_0^\infty dx\ x^{a_--1} (x+1)^{2-a_+} (2x+1)^{-1}
\sum_{k=0}^3 \gamma_k (x+1)^{-k}\ee
where
\be \gamma_0 = (a_+-a_-+3)(a_+-a_-+1)\ \ , \ \ \gamma_1 = -a_+ (3(a_+-a_-+1)+4)
\nonumber\ee
\be \gamma_2 = \frac{a_+((3a_++1)(a_+-a_-+1) +2a_+ )}{a_+-a_-+1}\ \ , \ \
\gamma_3 = -\frac{a_+^2(a_++1)}{a_+-a_-+1}\ee
After integrating, we obtain
\be \mathcal{B}_1 = \frac{1}{2(a_--1)(a_--2)}\
\sum_{k=0}^3 \gamma_k B(a_-,a_+-a_-+k-1)F(1,a_-;a_++k-1;-1)\ee
This is to be compared with $\mathcal{B}_0$ (eq.~(\ref{eq21})). Setting
\be\label{eqnew} c-a_- = -1+\epsilon\ee
we have
\be\label{eq56} \mathcal{B}_0 = - \frac{\Gamma(a_--2)\Gamma(1+a_+-a_-)}{\Gamma(a_+)}\ \epsilon +\dots
\ee
We may write
\be \mathcal{B}_1 = \frac{\Gamma(a_--2)\Gamma(1+a_+-a_-)}{2\Gamma(a_+)}\ \sum_{k=0}^3 \gamma_k' \ F(1,a_-;a_++k-1;-1)\ee
where
\be \gamma_k' = \gamma_k\ \frac{\Gamma(a_+)}{\Gamma(a_++k-1)}\
\frac{\Gamma(a_+-a_-+k-1)}{\Gamma(a_+-a_-+1)}\ee
Explicitly,
\be
\gamma_0' = (a_+-1)\left( 1+\frac{5}{a_+-a_-} +\frac{8}{(a_+-a_-)(a_+-a_--1)}\right)
\ \ , \ \ \gamma_1' = - a_+\left( 3+\frac{7}{a_+-a_-}\right)\ee
\be
\gamma_2' = 3a_++1+\frac{2a_+}{a_+-a_-+1}\ \ , \ \ \gamma_3' =-a_+\ee
Notice that all coefficients are $o(h_+)$, whereas we expect to finally get
an expression which is $o(1/h_+)$. This means that terms have to conspire
so that both $o(h_+)$ and $o(1)$ contributions cancel. In comparison, recall that
in the $n=1$ case, only $o(1)$ terms appeared and they were seen to cancel.

Let us expand the coefficients to the desired order,
\be \gamma_0' = a_++\frac{4a_++a_-}{a_+-a_-} +\frac{3a_++5a_-}{(a_+-a_-)^2}
+\dots \ \ , \ \  \gamma_1' = - 3a_+ - \frac{7a_+}{a_+-a_-}\ee
\be
\gamma_2' = 3a_++\frac{3a_+-a_-}{a_+-a_-}-\frac{2a_+}{(a_+-a_-)^2} +\dots \ \ , \ \ \gamma_3' =-a_+\ee
We may also use (\ref{eqhyp}) to expand the Hypergeometric functions (higher-order terms are not needed, because $\sum \gamma_k' = o(1/h_+)$). We obtain
\be F(1,a_-;a_++k-1;-1) = \frac{a_+}{a_++a_-} + \frac{a_-(ka_++(k-2)a_-)}{(a_++a_-)^3}
 + \dots\ee
It is now a straightforward exercise to show that all terms $o(h_+)$ and $o(1)$
cancel and we finally obtain
\bea\label{eq64} \mathcal{B}_1 &=& \frac{\Gamma(a_--2)\Gamma(1+a_+-a_-)}{2\Gamma(a_+)} \ \left\{ \frac{a_+^3+5a_+^2a_-+4a_+a_-^2+2a_-^3}{(a_+^2-a_-^2)^2} +\dots \right\} \nonumber\\
&=& \frac{\Gamma(a_--2)\Gamma(1+a_+-a_-)}{2\Gamma(a_+)} \ \left\{ \frac{12i-1}{16h_+} +\dots \right\}\eea
Using eqs.~(\ref{eq40}), (\ref{eq56}) and (\ref{eq64}), eq.~(\ref{eqnew}) reads
\be c-a_- = -1 + \frac{12i-1}{32h_+}\ee
The corresponding frequency is
\be\label{eqoe2} \hat\omega_2 = -2(1+i) \left\{ h_+ + \half + \frac{12i-1}{32h_+} +\dots \right\}\ee
For a general $n$ (eq.~(\ref{eqqnf})), $\mathcal{K}_+ = F_0$ contains a Polynomial of order $n-1$
in $(x+1)^{-1}$ (the Hypergeometric function in eq.~(\ref{eqsolinf})). Thus, the
complexity of the calculation increases with increasing $n$.
One obtains coefficients which are of order $h_+^{n-1}$. Even though they can be
seen to cancel order-by-order, leaving an expression of order $1/h_+$ after the dust settles,
yielding
\be\label{eqo1} \hat\omega_n = -2(1+ i)(n+h_+-{\textstyle{\frac{3}{2}}}) + o(1/h_+)\ \ , \ \ n=1,2,\dots\ee
the calculation of the integral~(\ref{eq39}) is uninspiring. We calculated
the coefficient of $1/h_+$ in~(\ref{eqo1}) explicitly in the cases $n=1$ (eq.~(\ref{eqoe1})) and $n=2$ (eq.~(\ref{eqoe2})).
It would be desirable to have an expression for the first-order correction~(\ref{eq39}) which is manifestly of order $1/h_+$.

It would also be interesting to extend the above results to more general black holes
in AdS$_d$ and compare with existing numerical results (see, e.g.,~\cite{bibw4}). The number of (unphysical) singularities increases if $d\ne 3,5$, so the analysis
is expected to be more complicated.
%

\end{document}